\documentclass[12pt]{article}
\usepackage{amsfonts,bm}
\usepackage{graphicx}
\baselineskip
\offinterlineskip
\begin{document}

\begin{center}
{\bf Majorana Fermions on a Lattice and a Matrix Problem}
\end{center}

\begin{center}
{\bf  Yorick Hardy$^\ast$ and Willi-Hans Steeb$^\dag$} \\[2ex]

$\ast$
Department of Mathematical Sciences, \\
University of South Africa, Pretoria, South Africa, \\
e-mail: {\tt hardyy@unisa.ac.za}\\[2ex]

$\dag$
International School for Scientific Computing, \\
University of Johannesburg, Auckland Park 2006, South Africa, \\
e-mail: {\tt steebwilli@gmail.com}\\[2ex]
\end{center}

\strut\hfill

{\bf Abstract} We investigate matrix problems 
arising in the study of Majorana Fermions on a lattice. 

\strut\hfill

Majorana Fermions on a lattice have been studied by many
authors \cite{1,2,3,4,5,6,7,8,9} in particular in connection
with Fermionic quantum computing. Here the following matrix
problem arises: Find a pair of $n \times n$ hermitian matrices
$A$ and $B$ such that $A^2=I_n$, $B^2=I_n$ and
$[A,B]_+=0_n$ (i.e. the anticommutator vanishes). Here $I_n$
is the $n \times n$ identity matrix and $0_n$ is the
$n \times n$ zero matrix. An example of such a pair of matrices
is given by
$$
A = \sigma_1 = \pmatrix { 0 & 1 \cr 1 & 0 }, \qquad
B = -\sigma_2 = \pmatrix { 0 & i \cr -i & 0 }
$$
where $\sigma_1$, $\sigma_2$, $\sigma_3$ are the Pauli spin matrices.
Note that the Pauli spin matrices satisfy $[\sigma_1,\sigma_2]_+=0_2$,
$[\sigma_2,\sigma_3]_+=0_2$, $[\sigma_2,\sigma_1]_+=0_2$ and
$\sigma_1^2=\sigma_2^2=\sigma_3^2=I_2$. 
\newline

Here we derive properties of such pairs of $n \times n$ matrices $A$ and $B$
and give some applications.
\newline  

First we show that the conditions cannot be satisfied if $n$ is odd. 
From $AB=-BA$ we obtain $\det(AB)=\det(-BA)$. Thus $\det(AB)=(-1)^n\det(AB)$.
Since $\det(AB) \ne 0$ this implies that the condition
$[A,B]_+=0_n$ cannot be satisfied if $n$ is odd.
\newline

Thus in the following we assume that $n$ is even
and $A$, $B$ satisfy the conditions  $A^2=I_n$, $B^2=I_n$ 
and $AB+BA=0_n$. From the conditions $A^2=I_n$, $B^2=I_n$ 
it follows that $A=A^{-1}$, $B=B^{-1}$. 
From the condition $AB+BA=0_n$ we also find that
$\mbox{tr}(AB)=0$ and therefore $\det(e^{AB})=1$.
It also follows that $(AB)^2=(BA)^2=-I_n$. 
\newline

Next we show that half of the eigenvalues of $A$ (and of $B$)
must be $+1$ and the other half must be $-1$.
Let $\mathbf{v}$ be an eigenvector of $A$ corresponding to the
eigenvalue 1. Then from
$$
(AB+BA)\mathbf{v}=AB\mathbf{v}+B\mathbf{v}=(A+I_n)B\mathbf{v}=
\mathbf{0}
$$
and the fact that $\mathbf{v}\neq \mathbf{0}$, 
$B\mathbf{v}\neq \mathbf{0}$ it follows
that $B\mathbf{v}$ is an eigenvector of $A$ corresponding to the
eigenvalue $-1$. Similarly if $\mathbf v$ is an eigenvector of $A$
corresponding to the eigenvalue $-1$, then $B\mathbf{v}$ is an eigenvector
of $A$ corresponding to the eigenvalue 1. Since $B$ is invertible
it follows that the eigenspaces of $A$ corresponding to the eigenvalues
1 and $-1$ have the same dimension.
\newline

The above argument is also constructive, i.e. given $A$ and an orthonormal basis
$$
\{\,\mathbf{v}_{-1,1},\,\mathbf{v}_{-1,2},\,\ldots,\mathbf{v}_{-1,n/2},
\,\mathbf{v}_{1,1},\,\mathbf{v}_{1,2},\,\ldots,
\mathbf{v}_{1,n/2}\,\}$$
composed of eigenvectors $\mathbf{v}_{-1,j}$ corresponding to the 
eigenvalue $-1$ of $A$ and eigenvectors $\mathbf{v}_{1,j}$ 
corresponding to the eigenvalue 1 of $A$,
we can construct a $B$ satisfying the properties above as
$$
B=\sum_{j=1}^{n/2}(\mathbf{v}_{1,j}
\mathbf{v}_{-1,j}^*+\mathbf{v}_{-1,j}\mathbf{v}_{1,j}^*).
$$
All $B$'s can be constructed in this way by an appropriate choice of
an orthonormal basis.
\newline

For any $n \times n$ matrices $X$ and $Y$ over $\mathbb C$ we have
the following expansion utilizing the anticommutator \cite{10,11}
$$
e^X Y e^X = Y + [X,Y]_+ + \frac1{2!} [X,[X,Y]_+]_+ +
\frac1{3!} [X,[X,[X,Y]_+]_+]_+ + \cdots\,.
$$ 
Consequently we find
$$
e^X Y e^{-X} = \left(Y + [X,Y]_+ + \frac1{2!} [X,[X,Y]_+]_+ +
\frac1{3!} [X,[X,[X,Y]_+]_+]_+ + \cdots\right)e^{-2X}
$$ 
$$
e^X Y e^{-X} = e^{2X}\left(Y - [X,Y]_+ + \frac1{2!} [X,[X,Y]_+]_+ -
\frac1{3!} [X,[X,[X,Y]_+]_+]_+ + \cdots\right)\,.
$$ 
Thus since $[A,B]_+=0_n$ for the matrices $A$ and $B$ we have
$$
e^A B e^{-A} = Be^{-2A} \quad \mbox{and} \quad
e^A B e^{-A} = e^{2A} B\,.
$$
Utilizing $B^2=I_n$ we have $e^{-A}=Be^{A}B$. Note that 
$(z \in {\mathbb C})$ 
$$
e^{zA} = I_n \cosh(z) + A\sinh(z), \quad
e^{zB} = I_n \cosh(z) + B\sinh(z)
$$
and 
$$
e^{zAB} = I_n \cos(z) + AB\sin(z)\,. 
$$
Let $\otimes$ be the Kronecker product \cite{12,13,14,15}.
Then we have
$$
e^{z(A \otimes B)} = (I_n \otimes I_n)\cosh(z) + (A \otimes B)\sinh(z)
$$
and for the anticommutators 
$$
[A \otimes I_n,B \otimes I_n]_+ = 0_{n^2}, \qquad 
[I_n \otimes A,I_n \otimes B]_+ = 0_{n^2}
$$
where $(A \otimes I_n)^2=I_n \otimes I_n$, 
$(B \otimes I_n)^2=I_n \otimes I_n$. Thus using the Kronecker
product and the identity matrix $I_n$ we can construct 
matrices in higher dimensions which satisfy the conditions.
This can be extended.
Let $C$ be an $n \times n$ matrix with $C^2=I_n$. Then
the pair $C \otimes A$ and $C \otimes B$ also satisfies
the conditions $(C \otimes A)^2=I_n \otimes I_n$,
$(C \otimes B)^2=I_n \otimes I_n$, 
$[C \otimes A,C \otimes B]_+=0_{n^2}$. An example for
$n=2$ would be $A=\sigma_3 \otimes \sigma_1$, 
$B=-\sigma_3 \otimes \sigma_2$. However note that 
\begin{eqnarray*}
[A \otimes A,B \otimes B]_+ &=& 2((AB) \otimes (AB)), \cr
[A \otimes B,A \otimes B]_+ &=& 2I_n \otimes I_n, \cr
[A \otimes B,B \otimes A]_+ &=& -2((AB) \otimes (AB))\,.
\end{eqnarray*}
Let $\oplus$ be the direct sum and $A$, $B$ be a pair of hermitian
matrices satisfying $A^2=B^2=I_n$ and $[A,B]_+=0_n$.
Then $A \oplus A$ and $B \oplus B$ is also such a pair (in $2n$ dimensions).
Let $A$, $B$ be such a pair with $n=2$. Then the $4 \times 4$ matrices 
$$
A \star A := 
\pmatrix { a_{11} & 0 & 0 & a_{12} \cr 0 & a_{11} & a_{12} & 0 \cr
           0 & a_{21} & a_{22} & 0 \cr a_{21} & 0 & 0 & a_{22} }
$$ 
and $B \star B$ are also such a pair.
\newline

From the pair $A$, $B$ we can also form the $n \times n$ matrix 
$A+iB$ which is non-normal, i.e. $(A+iB)^*(A+iB) \ne (A+iB)(A+iB)^*$. 
Such matrices play a role for the study of non-hermitian Hamilton 
operators \cite{16}. Since $(A+iB)^2=0_n$ we obtain
$$
e^{z(A+iB)} = I_n + z(A+iB)\,.
$$

\strut\hfill

{\bf Acknowledgment}
\newline

The authors are supported by the National Research Foundation (NRF),
South Africa. This work is based upon research supported by the National
Research Foundation. Any opinion, findings and conclusions or recommendations
expressed in this material are those of the author(s) and therefore the
NRF do not accept any liability in regard thereto.

\end{document}